\documentclass[fleqn,usenatbib, twocolumn]{mnras}
\usepackage[T1]{fontenc}
\usepackage{ae,aecompl}
\usepackage{rotating}
\usepackage{caption}
\usepackage{graphicx}
\usepackage{longtable}
\usepackage{xcolor}
\usepackage{ulem}
\usepackage{url}

\usepackage{graphicx}	
\usepackage{amsmath}	
\usepackage{amssymb}	
\usepackage{color}
\usepackage{url}
\usepackage{comment}

\title[]{Timing analysis of 2S 1417-624 observed with \textit{NICER} and \textit{Insight-HXMT}}

\author[L., Ji et al.]{%
	L. Ji$^{1}$\thanks{E-mail: ji.long@astro.uni-tuebingen.de},
	V. Doroshenko $^{1}$,
	A. Santangelo$^{1}$,
	C. {G{\"u}ng{\"o}r}$^{2}$,
	S. Zhang$^{3}$,
	L. Ducci  $^{1,4}$,
	\newauthor 
	S.-N. Zhang$^{3}$,
	M.-Y. Ge$^{3}$,
	L.J.\,Qu$^{3}$, Y.P.\,Chen$^{3}$, Q.C.Bu$^{3, 1}$
% ...
\\
	$^{1}$ Institut f\"ur Astronomie und Astrophysik, Kepler Center for Astro and Particle Physics, Eberhard Karls Universit\"at, Sand 1,\\ 72076 
	T\"ubingen, Germany\\
	$^{2}$Sabanc{\i} University, Faculty of Engineering and Natural Science, Orhanl{\i} $-$ Tuzla, 34956, \.{I}stanbul, Turkey\\
	$^{3}$Key Laboratory for Particle Astrophysics, Institute of High Energy Physics, Beijing 100049, China\\
	$^{4}$ ISDC Data Center for Astrophysics, Universit\'e de Gen\`eve, 16 chemin d'\'Ecogia, 1290 Versoix, Switzerland\\
	$^5$Department of Astronomy, Beijing Normal University, Beijing 100088, People’s Republic of China\\
	$^6$Department of Physics, Tsinghua University, Beijing 100084, People’s Republic of China\\
	$^7$University of Chinese Academy of Sciences, Chinese Academy of Sciences, Beijing 100049, People’s Republic of China\\
}

\date{Accepted XXX. Received YYY; in original form ZZZ}

\pubyear{2019}

\begin{document}
\label{firstpage}
\pagerange{\pageref{firstpage}--\pageref{lastpage}}
\maketitle

\begin{abstract}
We present a study of timing properties of the accreting pulsar 2S 1417-624 observed during its 2018 outburst, based on \textit{Swift}/BAT, \textit{Fermi}/GBM, \textit{Insight-HXMT} and \textit{NICER} observations.
We report a dramatic change of the pulse profiles with luminosity. 
The morphology of the profile in the range 0.2-10.0\,keV switches from double to triple peaks at $\sim2.5$ $\rm \times 10^{37}{\it D}_{10}^2\ erg\ s^{-1}$ and from triple to quadruple peaks at $\sim7$ $\rm \times 10^{37}{\it D}_{10}^2\ erg\ s^{-1}$.
The profile at high energies (25-100\,keV) shows significant evolutions as well.
We explain this phenomenon according to existing theoretical models.
We argue that the first change is related to the transition from the sub to the super-critical accretion regime, while the second to the transition of the accretion disc from the gas-dominated to the radiation pressure-dominated state. 
Considering the spin-up as well due to the accretion torque, this interpretation allows to estimate the magnetic field self-consistently at $\sim7\times 10^{12}$\,G.

\end{abstract}


\section{Introduction}
X-ray pulsars are highly magnetised neutron stars in binary systems with B $\sim 10^{12}$\,G, fed by accretion of matter from a donor star. Accretion is one of the most efficient mechanisms known to produce energy, and in fact the observed luminosity of these objects can reach $\sim10^{40}$\,$\rm erg\ s^{-1}$, making them natural 'laboratories' to study properties of matter under extreme conditions such as very high temperatures and ultra-strong magnetic fields.
Pulsations arise because plasma from the accretion disc is channelled by the magnetic field lines onto the magnetic poles, producing beamed radiation, which changes the orientation with respect to observers as the neutron star rotates.
Interaction of the accretion flow with the magnetic field thus defines the geometry of the emission region and plays a key role in X-ray pulsars \citep[e.g.,][]{Basko1976,Becker2007,Mushtukov2015c}. 

In particular, two regimes of accretion can be identified. 
If the radiation pressure can be ignored and the luminosity is smaller than a critical value $L_{\rm crit}$, i.e., in the sub-Eddington regime, the accreted plasma falls onto the surface of the neutron star forming accretion mounds. Otherwise, a radiation-dominated shock appears at some distance above the neutron star surface \citep[see, e.g., ][]{Basko1976, Becker2012,Mushtukov2015b}. The radiation pressure and consequently the transitional luminosity are defined by the geometry of the emission region, and thus by the magnetic field strength and structure. Observations of the transition between the two regimes and the study of the properties of X-ray pulsars in the two states is essential to understand the complex interplay between the ram pressure, the radiation structure and the intensity and geometry of the magnetic field.

2S 1417-624 is a transient source discovered with \textit{SAS-3} in 1978 \citep{Apparao1980}.
Several outbursts from the source have been observed by \textit{BATSE} and \textit{Rossi X-ray Timing Explorer (RXTE)} \citep{Finger1996, Inam2004, Gupta2018}.
The spin period was found to be $\sim$ 17.5\,s, while the spin-up rate was measured to be in the order of magnitude of $10^{-11}$\, ${\rm Hz\ s^{-1}}$, and correlated with the pulsed flux.
The optical counterpart has been identified as a B1 Ve star. 
Its orbital period of the source has been estimated by \citet{Finger1996} as $\sim$42\,days.
Using the optical properties of the donor star, \citet{Grindlay1984} estimated the distance of the binary system to be between 1.4 to 11.1\,kpc. Recently, \textit{Gaia} provided a distance measurement of $9.9_{-2.4}^{+3.1}$\,kpc (68\% confidence level) \footnote{source\_id=5854175187710795136} \citep{Bailer2018}.
Using \textit{Chandra} data, \citet{Tsygankov2017} detected a quiescence flux of $F_{(0.5-10\,keV)} \sim 5\times10^{-13}$\,erg\ s$^{-1}$ cm$^{-2}$, and modelled the \textit{Chandra} spectrum with a blackbody-like function with a temperature $\sim$1.5\,keV.
The derived high temperature suggests that in the low luminosity state the source might still accrete matter, without entering the propeller regime, in which centrifugal forces inhibit accretion \citep{Tsygankov2016,Gungor2017}.

In this paper, we report a study of timing properties of 2S 1417-624 during the outburst in 2018 using high cadence observations in a broad range of energy obtained with several facilities. 
This paper is organised as follows: the details of the observations and data reduction are introduced in Section 2; the results are presented in Section 3; and finally our  arguments to explain the observed phenomenology are discussed in Section 4.

\section{Data analysis and results}
The Burst Alert Telescope (BAT) onboard the \textit{Swift} observatory \citep{Gehrels2004} is an all-sky hard X-ray monitor aimed at studying transient phenomena such as gamma ray bursts. We have used the daily lightcurve (15-50\,keV) of the source provided by the BAT hard X-ray transient monitor \footnote{\url{https://swift.gsfc.nasa.gov/results/transients/weak/H1417-624/}} as an indicator of the bolometric flux during the outbursts. 
Another estimate of the source flux in the hard band is provided by \textit{Fermi}/Gamma-ray Burst Monitor (GBM) \footnote{\url{https://gammaray.nsstc.nasa.gov/gbm/science/pulsars/lightcurves/2s1417.html}} \citep{Meegan2009}. We have also used the spin-frequency and frequency derivative estimated by the GBM.
The pulsed flux reported is in the energy band of 12-50\,keV, which only includes the first and second harmonics.

\begin{figure}
    \centering
    \includegraphics[width=3.2in]{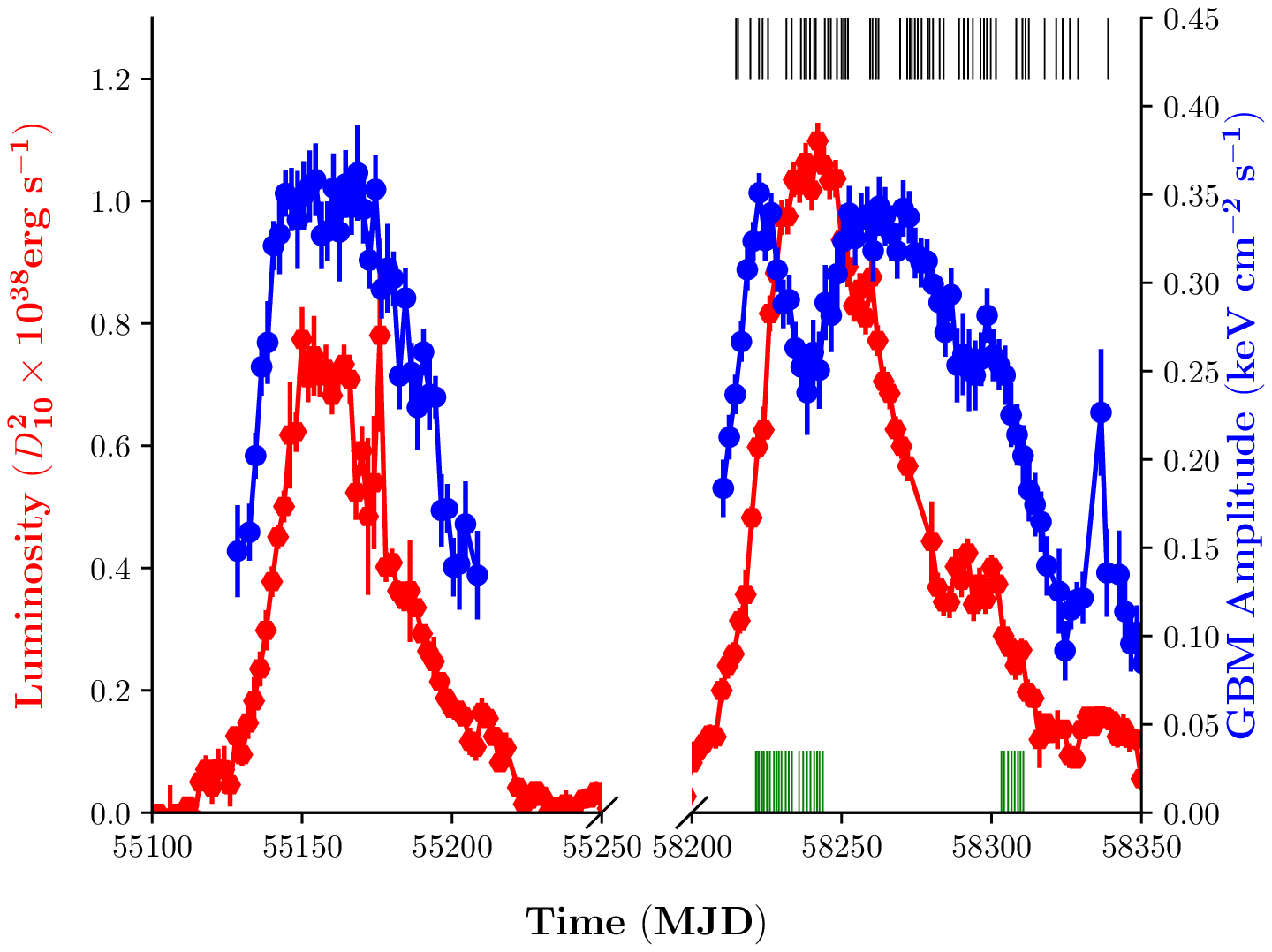}
    \includegraphics[width=3.2in]{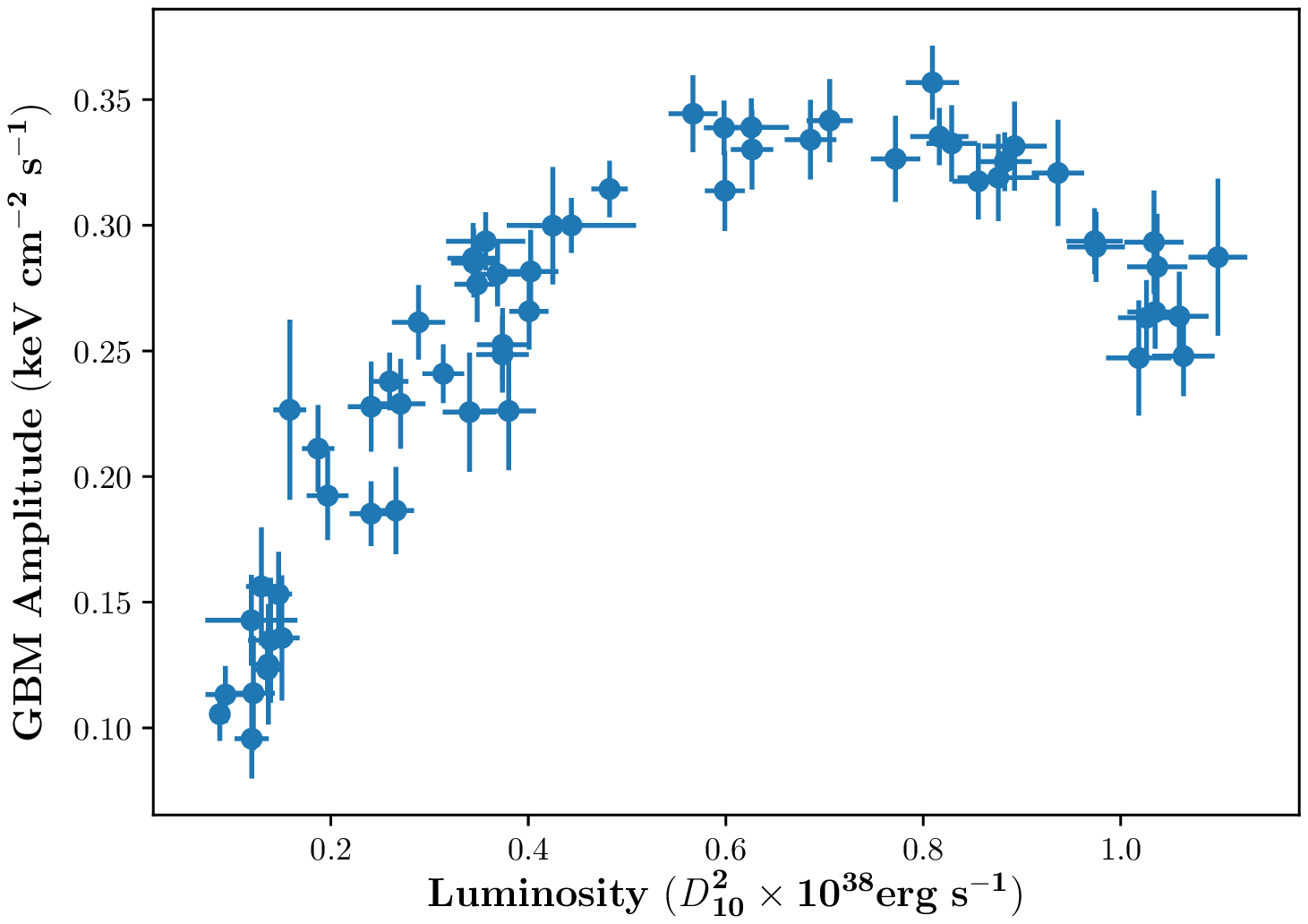}
    \includegraphics[width=3.2in]{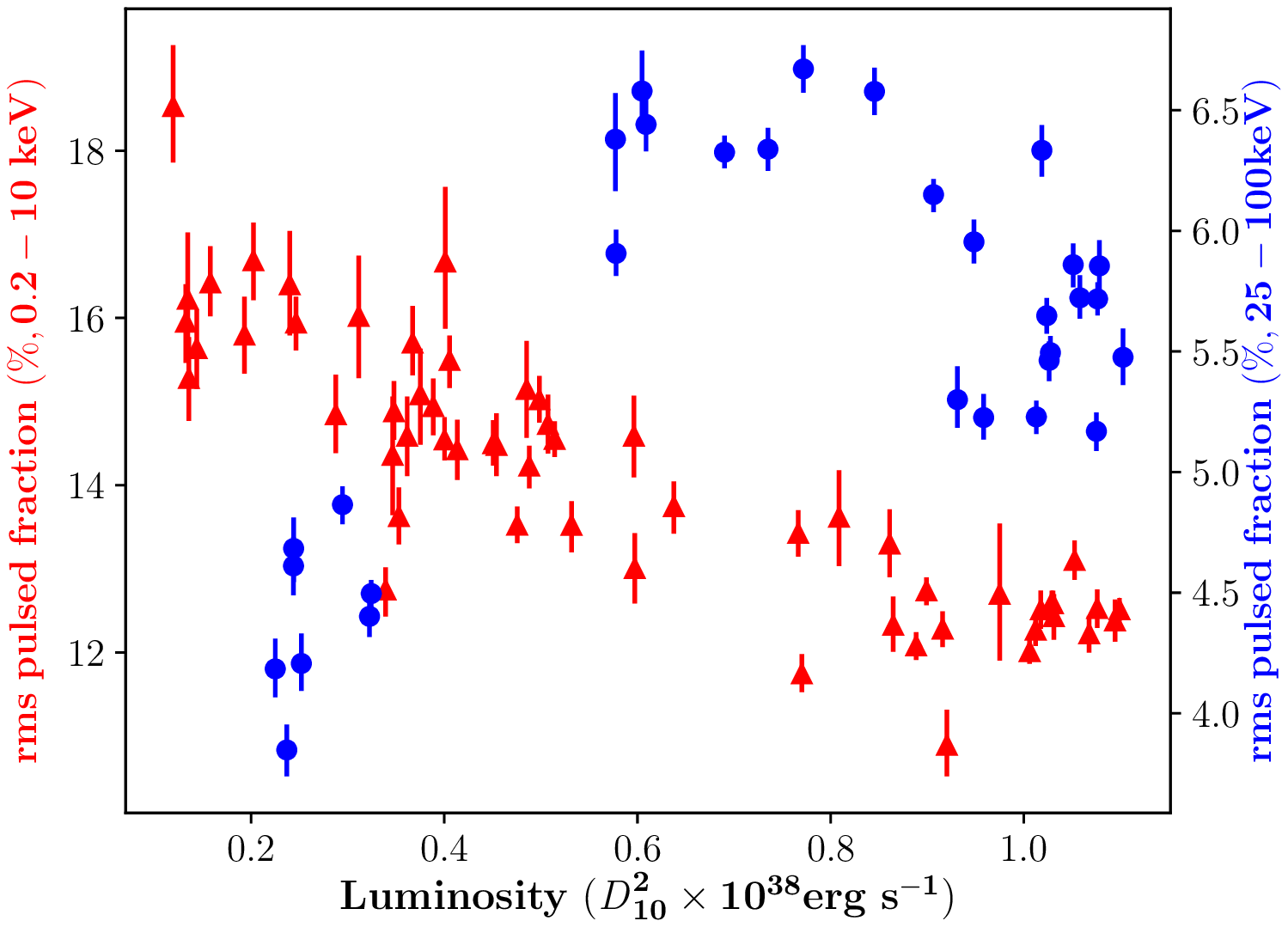}        

    \caption{
      Upper panel: 
      the luminosity of outbursts in 2009 and 2018
      observed with \textit{Swift}/BAT, after considering the bolometric correction performed by \textit{Insight-HXMT},
      and the pulsed flux observed with \textit{Fermi}/GBM (12-150\,keV).
      The black and green vertical lines represent the observational time of \textit{NICER} and \textit{Insight-HXMT}, respectively.
      Middle panel: the luminosity vs. the pulsed flux mentioned above for the 2018 outburst. 
       Bottom panel:
       The rms pulsed fraction observed with \textit{NICER} (red triangles; 0.2-10.0\,keV) and \textit{Insight-HXMT} (blue points; 25-100\,keV) for the outburst in 2018.
    }
    \label{lc}
\end{figure}

We have also used dedicated observations of the source obtained by the Hard X-ray Modulation Telescope (\textit{Insight-HMXT}) \citep{Zhang2014} in the hard band and the Neutron Star Interior Composition Explorer (\textit{NICER}) \citep{Gendreau2016} in the soft band.
China launched \textit{Insight-HMXT} in 2017. The mission has a wide energy coverage in the energy range 1-250\,keV and large effective area, especially at hard X-rays (>25\,keV).
There are 29 \textit{Insight-HMXT} pointing observations during the outburst of 2S 1417-624, in both the rising and decay phases.
Green vertical lines in the upper panel of Figure~\ref{lc} show observational times of \textit{Insight-HXMT}.
In this paper, we have used \textit{Insight-HXMT} data to estimate the bolometric flux, and the pulsed fraction at high energies (25-100\,keV). This allowed a cross check of the \textit{Fermi}/GBM's results.
The data analysis was performed with {\sc hxmtdas} v2.01 following the recommended  procedures in the user's guide \footnote{\url{http://www.hxmt.org/images/soft/HXMT\_User\_Manual.pdf}}. 
An extensive study of the \textit{Insight-HXMT} observations aimed at fully characterising the spectral-timing behaviour as a function of luminosity is in preparation \citep{Can2019}.
The estimated bolometric fluxes are well correlated with the Swift/BAT count rate in the 15-50\,keV range, which thus appears to be a good tracer of the bolometric flux.
The conversion factor (A) was calculated by using the broad band spectra of \textit{Insight-HXMT}. In fact the spectral shape of the source is relatively steady, and can be fitted as a cutoff powerlaw model, with a photon index of $\sim$ 0.25 and a cutoff at $\sim$ 16\,keV.
We conclude that the bolometric flux can be estimated by multiplying the observed \textit{Swift}/BAT count rate by the conversion factor A $\rm \approx 1.13 \times 10^{-7}\ erg/cts$.

\textit{NICER} is an external payload on-board the International Space Station \citep{Gendreau2016}.
In this work, we used the X-ray Timing Instrument (XTI), which operates in the range 0.2-10.0\,keV.
\textit{NICER} performed 83 observations during the outburst of 2S 1417-624 in 2018.
Among them we selected 57 observations (Table~\ref{table1}) with effective exposure longer than the 300\,s required to obtain meaningful estimate of the source pulse profiles in each observation.
Black vertical lines in the upper panel of Figure~\ref{lc} represent the observational times.
We followed the standard analysis procedures outlined in instruments' documentation and used {\sc heasoft} v6.25 to extract source lightcurves.

The first result was obtained through the simple comparison of the \textit{Swift}/BAT and \textit{Fermi}/GBM lightcurves of the 2009 and 2018 outbursts (see Figure~\ref{lc}).
The duration of the two outbursts is similar, e.g., $\sim$ 350\,days, but the second outburst is significantly brighter. 
We note the striking difference between the BAT and GBM lightcurves. While during the outburst in 2009 the pulsed flux measured by GBM is strongly correlated with the BAT rate, this is clearly not the case for the 2018 outburst (the middle panel of Figure~\ref{lc}). Here the pulsed flux increases with the BAT rate when the source is relatively faint ($\lesssim$5\, $\times$ $\rm 10^{37}$ $D_{10}^{2}$ $\rm \, erg\ s^{-1}$). 
However, for a luminosity higher than $\sim$ 8 $\times$ $\rm 10^{37}$ $D_{10}^{2}$ $\rm \, erg\ s^{-1}$, 
the correlation breaks, and the pulsed flux starts to drop while the bolometric flux traced by BAT increases. 
We note that the 2009 outburst also shows some sign of the saturation of the increasing GBM pulsed flux at the some flux.

\begin{figure}
    \centering
    \includegraphics[width=3.2in]{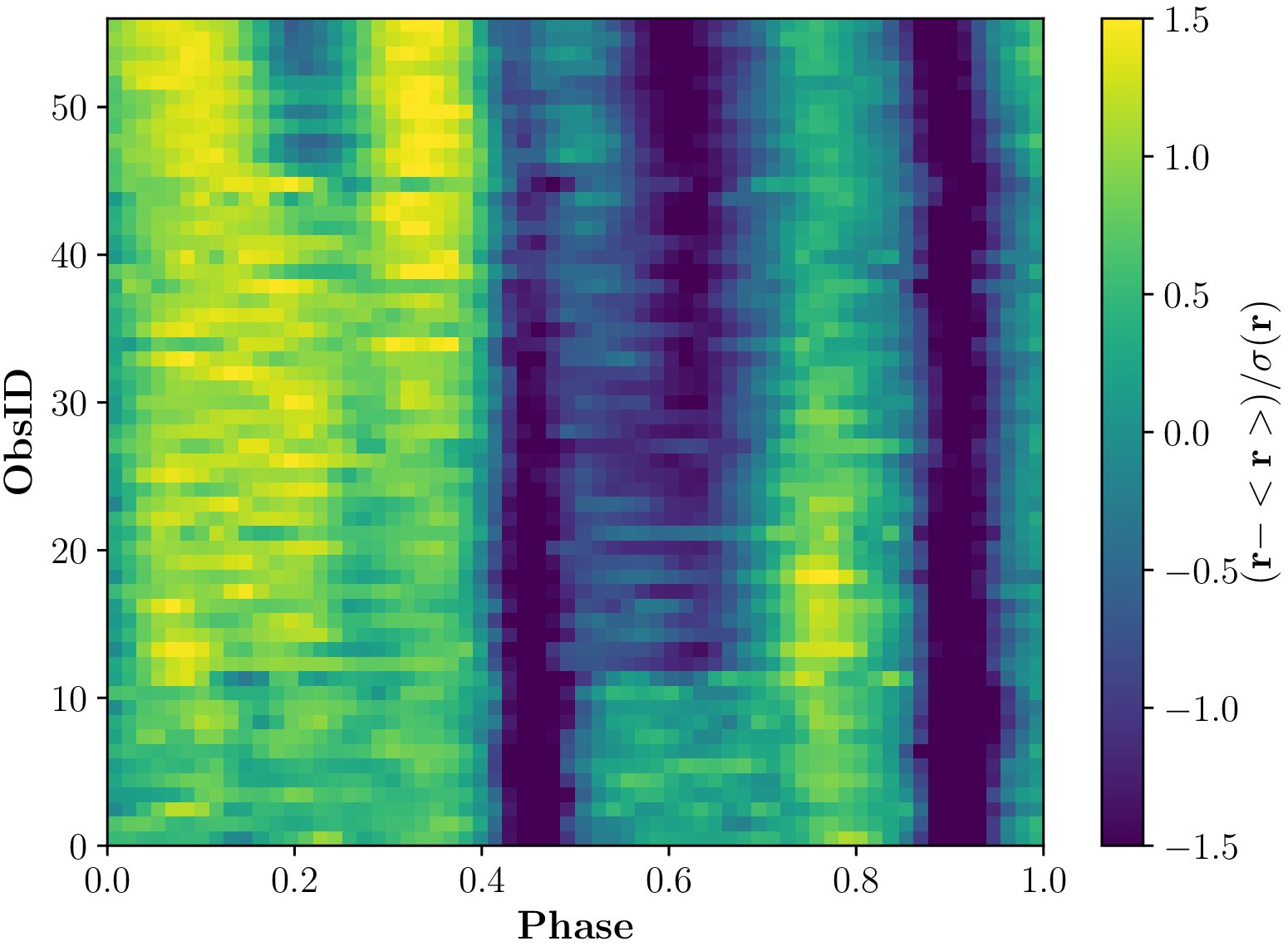}
    \includegraphics[width=3.2in]{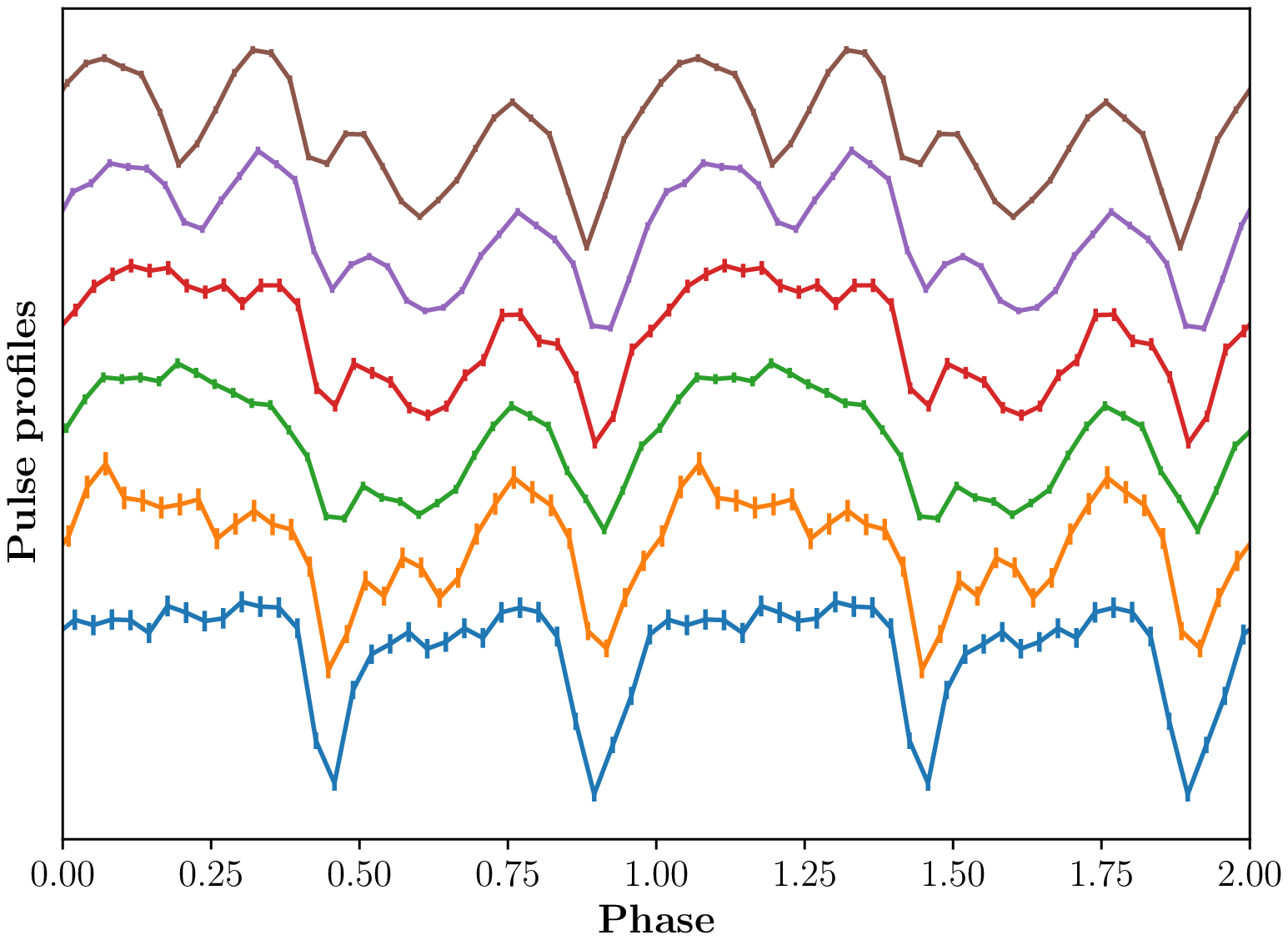}
    \caption{
        Upper panel:
        the evolution of pulse profiles of \textit{NICER} observations, where the ObsID is sorted (in an ascending order) according to the flux. The flux is estimated by using the \textit{Swift}/BAT count rate after taking into account the bolometric correction provided by \textit{Insight-HXMT}.
        Lower panel:
        representative pulse profiles at different flux levels (ObsID 6, 16, 26, 36, 46 and 56 from bottom to top).
    }
    \label{pulse_profile}
\end{figure}

This difference can only be due to a dramatic decrease of the pulsed fraction close to the peak of the second outburst. Such a dramatic change is also expected to affect the observed pulse profile shape.
To investigate that in more detail, we folded the events observed with \textit{NICER}/XTI based on the spin history and the orbital ephemeris reported by the \textit{Fermi}/GBM team.
We aligned the pulse profiles by cross-correlating pairs of the pulse profiles sorted by flux to obtain a "phase-luminosity" matrix shown in Figure~\ref{pulse_profile}.
Here the fluxes of the \textit{NICER} observations were estimated, using the contemporary BAT data and converting the observed BAT count rate to the bolometric flux as described above.

\begin{figure}
    \centering
    \includegraphics[width=3.2in]{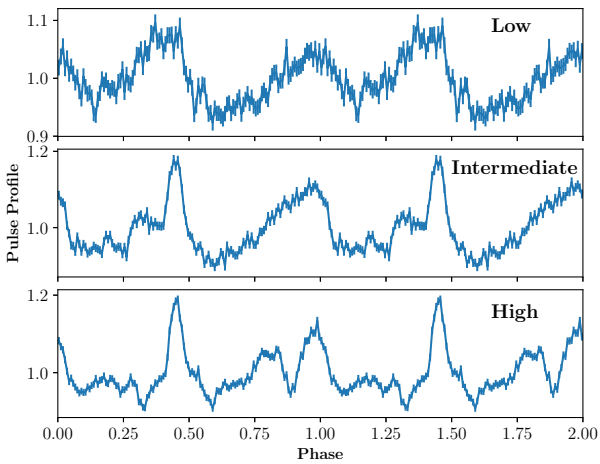}
    \caption{
    Examples of pulse profiles in the energy range of 25-100\,keV observed with \textit{Insight-HXMT}/HE in the low, intermediate and high states, respectively.
    }
    \label{pulse_profile_HXMT}
\end{figure}

Thanks to \textit{NICER}'s high cadence monitoring, the smooth evolution of the pulse profile morphology with luminosity can be observed. 
For a luminosity below $\sim$ 2.5 $\times$ $\rm 10^{37} {\it D_{\rm 10}^2} \, erg\ s^{-1}$ (obsID $\sim$ 10) the pulse profile at 0.2-10.0\,keV shows two broad peaks.
At a higher luminosity an additional peak appears, and a second transition of the morphology to even four peaks is observed when the luminosity is larger than $\sim$ 7 $\times$ $\rm 10^{37} {\it D_{\rm 10}^2} \, erg\ s^{-1}$.
For hard X-rays (25-100\,keV) observed with \textit{Insight}-HXMT/HE, the pulse profile also exhibits significant changes with luminosity.
For example, in Figure~\ref{pulse_profile_HXMT} we show pulse profiles at low, intermediate and high states (3,7,10$\times$ $\rm 10^{37} {\it D_{\rm 10}^2} \, erg\ s^{-1}$), respectively.
The pulse profile has two broad peaks at the low state, and one of them evolves into a narrower structure at a higher luminosity.
When the luminosity is larger than $\sim$ 7 $\times$ $\rm 10^{37} {\it D_{\rm 10}^2} \, erg\ s^{-1}$, a triple-peak profile is gradually shown.
We note that changes of the pulse profiles observed with \textit{Insight-HXMT} correspond to the variability of the GBM pulsed flux shown in Figure~\ref{lc}. 

We show the rms pulsed fractions (PF) for both soft (0.2-10.0\,keV) and hard (25-100\,keV) X-rays observed with \textit{NICER} and \textit{Insight-HXMT} in Figure~\ref{lc}.
The pulsed fraction is calculated as $\sqrt{\Sigma_{j=1}^{m} (a_{\rm j}^2 + b_{\rm j}^2 - \sigma_{\rm a, j}^2 - \sigma_{\rm b, j}^2) /  (a_{0}^2 + b_{0}^2 )} $, where $a_{\rm j}$ and $b_{\rm j}$ are the Fourier coefficient, $\sigma_{\rm a, j}$ and  $\sigma_{\rm b, j}$ are the corresponding uncertainties, and $m$ is the number of phase bins \citep{Archibald2015}.
The pulsed fraction both in the hard (25-100\,keV) and soft (0.2-10.0\,keV) bands
appears to change with luminosity, however, the dependence is different. In the soft band the pulsed fraction decreases with luminosity, whereas in hard X-rays it actually increases with the luminosity up to 7$\times$ $\rm 10^{37} {\it D_{\rm 10}^2} \, erg\ s^{-1}$, and then  decreases.
The pulsed fraction luminosity dependence in the hard band revealed by \textit{Insight-HXMT} thus confirms the already noted drop of the pulsed flux at the peak of the second outburst based on the comparison of Fermi/GBM and BAT fluxes.
It is also interesting to note that the drop of the pulsed fraction in the hard band occurred simultaneously with the transition of the soft X-ray pulse profiles from three peaked to four peaked shape. 
\section{Discussion}
The observed evolution of the pulse profiles with luminosity in both soft (0.2-10.0\,keV) and hard (25-100\,keV) energy bands suggests that two regime transitions occurs in the source: the first at 2.5$\times$ $\rm 10^{37} {\it D_{\rm 10}^2} \, erg\ s^{-1}$ ($L_{\rm crit}$), and the second at 7$\times$ $\rm 10^{37} {\it D_{\rm 10}^2} \, erg\ s^{-1}$ ($L_{\rm ZoneA}$).
We note that the first transition has been reported by \citet{Gupta2018} at a similar flux level, with \textit{RXTE} observations of the giant outburst in 2009.
Based on the observed spectral evolution of the source, they interpreted the first observational transition as due to changes of the pulsed beam when the pulsar goes from the sub-critical to the super-critical regime.
The second transition has not been reported previously. Most likely it did not occur in the previous outburst as it appears to be associated with a decrease of the pulsed fraction at high fluxes, which indeed was not been observed in the 2009 outburst. 
The origin of the second transition is poorly known. 
Here we suggest that it might be caused by the transition between the gas and radiation pressure dominated states of the inner regions of the accretion disc theoretically predicted by \citet{Shakura1973, Mushtukov2015} and recently discovered in Swift J0243.6+6124 by \citet{Doroshenko2019}. 
At high accretion rates the disc extends deeper into the magnetosphere of the neutron star, so that the temperature, energy release rate, and radiation pressure become sufficiently high to dramatically affect its structure within the disc.
In particular, the disc thickness increases, which affects the geometry of the accretion flow and thus the emission region geometry, the beam, and eventually the observed pulse profile shape \citep{Doroshenko2019}.
Differently than in the case of Swift J0243.6+6124, however, in 2S 1417-624 we were not able to detect significant changes in the power spectrum. 
In fact, the power spectrum in 2S 1417-624 appears to be consistent with a single power law throughout the outburst, and no breaks are observed at all.
We note, however, that origin of the observed breaks in the power spectrum of Swift J0243.6+6124 is not well known, and dramatically different power spectral shapes have been reported for different sources \citep{Monkkonen2019}.
The detailed study of the properties of the radiation pressure dominated (RPD) disc is out of scope of this paper and is complicated by the relatively poor statistics. Here we would only like to note the consistency of the proposed interpretation in the framework of existing theoretical estimates for critical luminosity and the formation of an RPD disc.
If our interpretation of both transitions is correct, we could constrain the distance and the magnetic field of the source, based on the following equations \citep{Becker2012, Andersson2005, Monkkonen2019}:
\begin{equation}
 \begin{array}{l}
 \begin{aligned}
L_{\rm Crit} 
 &= 1.49 \times 10^{37} \omega^{28/15} m^{29/30} R^{1/10}_6 B^{16/15}_{12} {\rm erg\ s^{-1}}\\
    L_{\rm ZoneA}\
    &= 3 \times 10^{38} k^{21/22} \alpha^{-1/11} m^{6/11} R^{7/11}_6 B_{12} ^ {6/11} {\rm erg\ s^{-1}}
 \end{aligned}
\end{array}
\end{equation}
Where $D$ is the distance, $m$, $R_{6}$ and $B_{12}$ are the mass, the radius and the magnetic field of the source in the units of 1.4$M_{\odot}$, $10^6$\,cm and $10^{12}$\,G, respectively.
We assume $\omega$ = 1 and $\alpha$ = 0.1 for typical parameters. 
The $k$ is a model-dependent dimensionless number between the magnetic radius and the Alfv$\rm \acute{e}$n radius, usually assumed to be $k$$\sim$ 0.5 \citep{Ghosh1979, Wang1996}.
Clearly, for a given $k$, the distance ($D$) and the magnetic field ($B$) can be determined.

\begin{figure}
    \centering
    \includegraphics[width=3.2in]{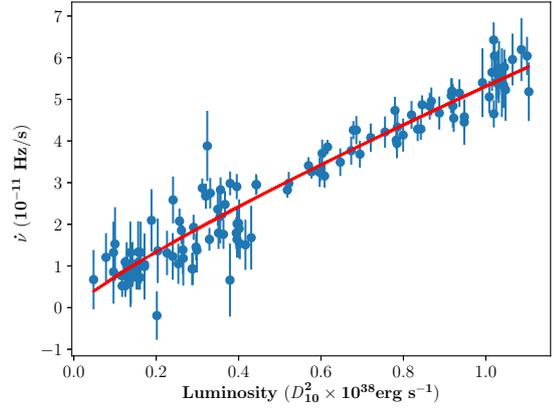}
    \\[15pt]
     \includegraphics[width=3.2in]{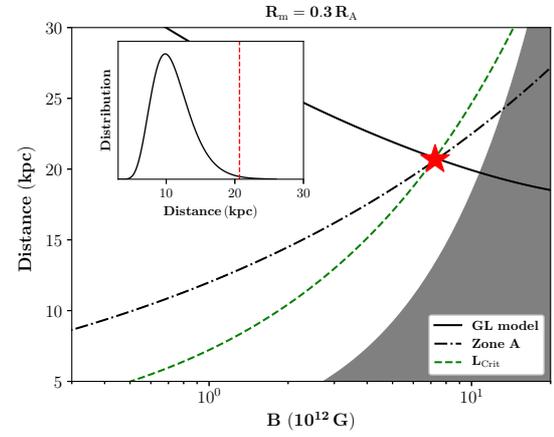}
    \caption{
        Upper panel: the luminosity vs. the frequency derivative, where the red line is the fitting by using GL model.
        Lower panel: the estimation of the magnetic field and distance of 2S 1417-624 (shown as a red star).
        The solid black line shows the fitting result by using the GL model.
        The green dashed line is obtained by  assuming $\rm F_{\rm crit1}$ is the critical luminosity between accretion regimes.
        The dash-dot line shows the condition if $\rm F_{\rm crit2}$ corresponds to the changes of the accretion disc between the gas and radiation pressure dominated states.
        The grey region represents the forbidden parameter space obtained from the propeller effect, only above which the source is able to accrete in the quiescence state as observed by  \citet{Tsygankov2017}. 
        The inset shows the distribution of the distance  suggested by \textit{Gaia} \citep{Bailer2018}, where the vertical line is the estimation in this work.
}
    \label{DB}
\end{figure}

The distance and the magnetic field can also be constrained by the observed spin-up rate of the pulsar induced by the accretion torque.  Using the model proposed by \citet{Ghosh1979} (GL model), we can estimate the spin-up rate with:
\begin{equation}
\dot{\nu} = 2^{-15/14}k^{1/2}{\mu}^{2/7}(GM)^{-3/7}(I\pi)^{-1}R^{6/7}{L}^{6/7}n(\omega)\ {\rm Hz\ s^{-1}}
\end{equation}
where $I=\frac{2}{5}MR^2$, $\mu=\frac{1}{2}BR^3$ and $F$ are the moment of inertia, the magnetic dipole moment and the bolometric flux, respectively.
$\omega$ is the fastness parameter, and n($\omega$)$\approx$1.4 for a slow rotator.
We note that the model depends on the parameter $k$.
We find that $k$ $\sim$ 0.3 is required for the above three equations to yield a consistent solution,  i.e., converge into one point in the $D$-$B$ diagram.
This implies $D$ and $B$ of $\sim$ $7\times 10^{12}$\,G and $\sim$ $20$\,kpc, respectively.
We show the fitting of the torque model and the resulting $D$-$B$ relation in Figure~\ref{DB}.
We note that \citet{Doroshenko2019} obtained a similar conclusion ($k$$\sim$0.25) for Swift J0243.6+6124, using the same method.

Another estimate of the field can be obtained by the fact that
the source likely continued to accrete in quiescence without switching to the propeller phase \citep{Tsygankov2017}.
The accretion luminosity in this case should be larger than  \citep{Campana2002, Tsygankov2016}:
\begin{equation}
 \begin{array}{l}
 \begin{aligned}
L > L_{\rm prop} &\approx 4\times 10^{37} k ^{7/2} B_{12}^2 P^{-7/3} m^{-2/3}R_6^5\ {\rm erg\ s^{-1}}
 \end{aligned}
\end{array}
\end{equation}

This condition is shown as a dotted line in in Figure~\ref{DB}.
The distance and magnetic field inferred above are consistent with the parameter space where the accretion is allowed in the quiescence state.

No cyclotron resonance scattering features (CRSFs) have been found in 2S 1417-624 with \textit{RXTE} observations, and in the preliminary spectral analysis of \textit{NuSTAR} as well.
The high value of the estimates of the magnetic field obtained above $\sim$ $7\times 10^{12}$\,G
could explain the lack of detection as the CRSF could be expected to have energy $\gtrsim$ 80\,keV in this case. Unfortunately, the counting statistics does not allow put robust detection of an absorption line at these energies \citep{Can2019}.
The inferred distance is $\sim$ 20\,kpc, which is however larger than the \textit{Gaia}'s estimation at a nearly 3\,$\sigma$ significance level.
We note that this discrepancy mainly originates from the torque model, which does not allow a closer distance.
Other torque models cannot solve this problem either because they predict a similar behaviour for slow rotators, like the case in 2S 1417-624 \citep[see, e.g.,][]{Wang1987, Kluzniak2007, Shi2015}.
If the distance measured by \textit{Gaia} is correct, a torque which is $\sim$ 3 times larger than GL model is required to explain the observed spin-up.
This may bring a challenge for the current torque models, and other effects, e.g., the quadrupolar magnetic field, might be important.
Furthermore, the $L_{\rm crit}$ is also highly uncertain \citep{Becker2012, Mushtukov2015c}, and its effect is discussed by \citet{Doroshenko2019}.
Nevertheless, a deep study that contains both the spin-up rate and the variability of pulse profiles provides a new measure and a self-consistent solution to understand the magnetic field of 2S 1417-624. 
On the other hand, independent estimate of the magnetic field, for instance by detection of a cyclotron line, would allow to verify theoretical assumptions we used above.
Observing similar phenomenology in more sources is also required to ensure the robustness of these interpretations, particularly in  ultraluminous X-ray sources.
\section*{Acknowledgements}
This work made use of the data from the Insight-HXMT mission, a project funded by China National Space Administration (CNSA) and the Chinese Academy of Sciences (CAS). The Insight-HXMT team gratefully acknowledges the support from the National Program on Key Research and Development Project (Grant No. 2016YFA0400800) from the Minister of Science and Technology of China (MOST) and the Strategic Priority Research Program of the Chinese Academy of Sciences (Grant No. XDB23040400). The authors thank supports from the National Natural Science Foundation of China under Grants No. 11503027, 11673023, 11733009,
U1838201 and U1838202.
 We acknowledge the use of public data and products from the \textit{Swift}, \textit{NICER} and \textit{Fermi} data archive.
\bibliographystyle{mnras}
\bibliography{mybibtex}

B
\appendix
\begin{table*}
    \small
    \caption{The columns denote numbers, ObsIDs, time, luminosity and pulsed fractions at 0.2-10.0\,keV of \textit{NICER} observations for the outburst in 2018.}
    \begin{tabular}{ccccc}
        \hline
        No. & ObsID      & Time     &   Luminosity           &      PF                \\                
        &            & (MJD)    & ($\rm {\it D}_{10}^2 \times 10^{38} erg\ s^{-1}$)  &      (\%)                \\                
        \hline
1  & 1200130177 & 58328.83 & $0.12 \pm 0.02$  & $18.59_{-0.69}^{+0.68}$\\ 
2  & 1200130175 & 58326.13 & $0.13 \pm 0.02$  & $15.96_{-0.47}^{+0.44}$\\ 
3  & 1200130173 & 58323.69 & $0.13 \pm 0.02$  & $16.13_{-0.77}^{+0.86}$\\ 
4  & 1200130171 & 58321.56 & $0.14 \pm 0.03$  & $15.31_{-0.48}^{+0.46}$\\ 
5  & 1200130169 & 58317.70 & $0.14 \pm 0.01$  & $15.63_{-0.48}^{+0.49}$\\ 
6  & 1200130181 & 58338.84 & $0.16 \pm 0.02$  & $16.45_{-0.42}^{+0.41}$\\ 
7  & 1200130167 & 58311.34 & $0.19 \pm 0.02$  & $15.79_{-0.49}^{+0.42}$\\ 
8  & 1200130168 & 58312.44 & $0.20 \pm 0.02$  & $16.65_{-0.45}^{+0.52}$\\ 
9  & 1200130166 & 58310.32 & $0.24 \pm 0.02$  & $16.39_{-0.57}^{+0.62}$\\ 
10  & 1200130165 & 58308.25 & $0.25 \pm 0.02$  & $15.95_{-0.35}^{+0.31}$\\ 
11  & 1200130104 & 58214.77 & $0.29 \pm 0.02$  & $14.90_{-0.45}^{+0.44}$\\ 
12  & 1200130105 & 58215.45 & $0.31 \pm 0.02$  & $15.97_{-0.78}^{+0.70}$\\ 
13  & 1200130149 & 58282.67 & $0.34 \pm 0.02$  & $12.75_{-0.29}^{+0.27}$\\ 
14  & 1200130154 & 58293.76 & $0.35 \pm 0.02$  & $14.40_{-0.70}^{+0.72}$\\ 
15  & 1200130156 & 58297.46 & $0.35 \pm 0.02$  & $14.84_{-0.34}^{+0.33}$\\ 
16  & 1200130155 & 58296.28 & $0.36 \pm 0.03$  & $13.61_{-0.36}^{+0.36}$\\ 
17  & 1200130157 & 58298.46 & $0.36 \pm 0.02$  & $14.55_{-0.51}^{+0.51}$\\ 
18  & 1200130150 & 58284.00 & $0.37 \pm 0.03$  & $15.72_{-0.42}^{+0.43}$\\ 
19  & 1200130160 & 58301.48 & $0.37 \pm 0.03$  & $15.06_{-0.58}^{+0.58}$\\ 
20  & 1200130151 & 58289.16 & $0.39 \pm 0.02$  & $14.94_{-0.35}^{+0.34}$\\ 
21  & 1200130158 & 58299.72 & $0.40 \pm 0.02$  & $16.67_{-0.81}^{+0.80}$\\ 
22  & 1200130148 & 58280.49 & $0.40 \pm 0.04$  & $14.55_{-0.26}^{+0.28}$\\ 
23  & 1200130152 & 58290.71 & $0.41 \pm 0.02$  & $15.48_{-0.31}^{+0.30}$\\ 
24  & 1200130153 & 58292.16 & $0.41 \pm 0.03$  & $14.40_{-0.37}^{+0.34}$\\ 
25  & 1200130147 & 58279.27 & $0.45 \pm 0.05$  & $14.49_{-0.30}^{+0.27}$\\ 
26  & 1200130146 & 58278.66 & $0.46 \pm 0.05$  & $14.48_{-0.41}^{+0.40}$\\ 
27  & 1200130145 & 58276.61 & $0.48 \pm 0.04$  & $13.52_{-0.19}^{+0.21}$\\ 
28  & 1200130106 & 58219.56 & $0.49 \pm 0.02$  & $15.14_{-0.63}^{+0.58}$\\ 
29  & 1200130144 & 58275.38 & $0.49 \pm 0.03$  & $14.21_{-0.26}^{+0.27}$\\ 
30  & 1200130143 & 58274.42 & $0.50 \pm 0.03$  & $15.05_{-0.29}^{+0.29}$\\ 
31  & 1200130142 & 58273.39 & $0.51 \pm 0.02$  & $14.72_{-0.36}^{+0.34}$\\ 
32  & 1200130141 & 58272.75 & $0.52 \pm 0.02$  & $14.57_{-0.23}^{+0.21}$\\ 
33  & 1200130140 & 58271.88 & $0.53 \pm 0.02$  & $13.52_{-0.29}^{+0.27}$\\ 
34  & 1200130107 & 58222.43 & $0.60 \pm 0.03$  & $13.01_{-0.46}^{+0.44}$\\ 
35  & 1200130139 & 58269.48 & $0.60 \pm 0.02$  & $14.56_{-0.48}^{+0.47}$\\ 
36  & 1200130108 & 58223.63 & $0.64 \pm 0.04$  & $13.71_{-0.32}^{+0.35}$\\ 
37  & 1200130135 & 58262.28 & $0.77 \pm 0.03$  & $13.42_{-0.29}^{+0.25}$\\ 
38  & 1200130134 & 58261.46 & $0.77 \pm 0.02$  & $11.75_{-0.22}^{+0.21}$\\ 
39  & 1200130110 & 58225.43 & $0.81 \pm 0.03$  & $13.58_{-0.57}^{+0.61}$\\ 
40  & 1200130132 & 58259.44 & $0.86 \pm 0.04$  & $13.31_{-0.37}^{+0.39}$\\ 
41  & 1200130130 & 58252.13 & $0.87 \pm 0.04$  & $12.35_{-0.35}^{+0.32}$\\ 
42  & 1200130133 & 58260.30 & $0.89 \pm 0.05$  & $12.07_{-0.16}^{+0.17}$\\ 
43  & 1200130129 & 58251.31 & $0.90 \pm 0.04$  & $12.73_{-0.18}^{+0.17}$\\ 
44  & 1200130128 & 58250.74 & $0.92 \pm 0.04$  & $12.27_{-0.20}^{+0.24}$\\ 
45  & 1200130127 & 58249.95 & $0.92 \pm 0.03$  & $10.87_{-0.38}^{+0.39}$\\ 
46  & 1200130113 & 58231.53 & $0.98 \pm 0.03$  & $12.68_{-0.83}^{+0.83}$\\ 
47  & 1200130126 & 58248.40 & $1.01 \pm 0.03$  & $12.01_{-0.15}^{+0.15}$\\ 
48  & 1200130118 & 58239.44 & $1.01 \pm 0.03$  & $12.28_{-0.19}^{+0.18}$\\ 
49  & 1200130115 & 58236.38 & $1.02 \pm 0.03$  & $12.51_{-0.24}^{+0.24}$\\ 
50  & 1200130124 & 58246.34 & $1.03 \pm 0.03$  & $12.58_{-0.14}^{+0.15}$\\ 
51  & 1200130114 & 58233.31 & $1.03 \pm 0.03$  & $12.59_{-0.14}^{+0.14}$\\ 
52  & 1200130123 & 58245.47 & $1.03 \pm 0.03$  & $12.43_{-0.28}^{+0.27}$\\ 
53  & 1200130122 & 58244.34 & $1.05 \pm 0.03$  & $13.10_{-0.22}^{+0.24}$\\ 
54  & 1200130116 & 58237.60 & $1.07 \pm 0.03$  & $12.21_{-0.22}^{+0.22}$\\ 
55  & 1200130117 & 58238.22 & $1.08 \pm 0.04$  & $12.52_{-0.23}^{+0.24}$\\ 
56  & 1200130119 & 58240.86 & $1.10 \pm 0.04$  & $12.37_{-0.24}^{+0.26}$\\ 
57  & 1200130120 & 58241.36 & $1.10 \pm 0.03$  & $12.50_{-0.14}^{+0.13}$\\ 
        \hline
    \end{tabular}
\label{table1}
\end{table*}

\end{document}